\documentclass[]{tMOP2e}
\begin{document}
\doi{10.1080/0950034YYxxxxxxxx}
 \issn{}
\issnp{} \jvol{} \jnum{} \jyear{2011} \jmonth{May}

\markboth{A. Wickenbrock, Piyaphat Phoonthong and Ferruccio Renzoni }{Journal of Modern Optics}


\title{Collective strong coupling in a lossy optical cavity}

\author{Arne Wickenbrock$^{\ast}$\thanks{$^\ast$Corresponding author. Email: a.wickenbrock@ucl.ac.uk
\vspace{6pt}},Piyaphat Phoonthong and Ferruccio Renzoni\\\vspace{6pt}  {\em{Department of Physics and Astronomy, University College London, London,
UK}}\\\vspace{6pt}\received{April 2011} }

\maketitle

\begin{abstract}
We observe vacuum Rabi splitting in a lossy nearly confocal cavity indicating 
the strong coupling regime, despite a weak single-atom single-mode coupling. Strong collective interaction manifests itself 
in the typical $\sqrt{N}$-dependence of the normal mode splitting on the 
number of atoms $N$.  
The $TEM_{00}$-mode coupling parameters are $\left( g,\kappa,\gamma\right)=2\pi\times\left(0.12,0.8,2.6\right)$ MHz and up to $\left(1.33\pm 0.08\right)\times10^5$ cesium atoms were loaded into the mode volume.\bigskip

\bigskip

\end{abstract}
\vspace{-42pt}
\section*{}

\section{Introduction}
The radiative properties of an atom are a function of the available 
electromagnetic 
vacuum modes, hence they are altered by the presence of an optical resonator. 
Correspondingly, spontaneous emission of an atom can be enhanced and suppressed by placing it 
inside a cavity depending on the cavity resonance - atom transition detuning 
as has been shown in \cite{Kleppner1}. For the last 30 years cavity quantum 
electrodynamics has been attracting attention as an ideal framework to explore
the foundations of quantum mechanics, but also with wide 
ranging application in metrology (e.g \cite{Kasevich2006}), quantum computing 
(e.g \cite{Zoller1}), quantum cryptography and the development of novel cavity 
assisted laser cooling mechanisms \cite{Domokos2003, Vuletic2001}.\\
In the far-off-resonance case, the interaction of an atom, or more generally a polarizable particle,
with the optical resonator can be understood classically as a position 
dependent refractive index changing the optical pathlength of the cavity 
and therefore its scattering and emission properties. The interaction strength 
between atom and cavity depends on the dipole moment $\mu$ of the scatterer 
compared to the overall mode volume $V_{mode}$ of the resonator, and the 
coupling constant is 
$g=\mu\sqrt{\omega_A/\left(2 \hbar \epsilon_0 V_{mode}\right)}$, where $\omega_A$ here is the frequency of the atomic transition.\\ 
Quantum mechanically the single-atom single-mode interaction is best described by the 
Jaynes-Cummings Hamiltonian \cite{Jaynes63}. In the new dressed basis the 
excited state is split into two components due to the vacuum fluctuation of the 
electro-magnetic field. The so called vacuum Rabi splitting is proportional to
the atom-cavity coupling constant and is observable when it is bigger than 
the cavity linewidth $\nu_C$ and the atomic transition linewidth $\Gamma$. The cavity-atom interaction can also be characterized by the 
so-called cooperativity parameter, i.e. the ratio of the photon scattering/emission rate into the cavity mode $\Gamma_C$ with respect to the free space scattering/emission rate $\Gamma_{FS}$. For a single atom - single mode interaction, the 
cooperativity parameter is given by:
	\[C=\frac{\Gamma_C}{\Gamma_{FS}}=\frac{g^2}{2 \kappa \gamma}~. \]
Here $\kappa=\nu_C/2$ is half the cavity loss rate and $\gamma=\Gamma/2$ is half the excited state decay rate.
If $g>>\left(\kappa+\gamma\right)$ and therefore $C>>1$ the coherent cavity-atom excitation exchange is much quicker than the dissipative 
decay into free space or the decay of the cavity mode. Therefore an emitted photon stays 
inside the cavity, eventually gets reabsorbed by the atom, reemitted into the cavity and so on. 
Finally, the photon will leave the system either into free space or through 
one of the mirrors. This regime is called the strong coupling regime.
Depending on the specific application, strong coupling may be a requirement. For example in the context of cavity cooling it was pointed out in Ref. \cite{Prospects1} that strong coupling is not essential for cavity-assisted laser cooling of two-level atoms, but it is important for cavity-cooling of multi-level atoms and molecules. There are now two ways to reach the strong coupling regime. One way is to reduce the mode-volume of the cavity and improve coating of the mirrors to produce a very large coupling constant $g$ 
and small cavity loss rate $\kappa$ as done for example in Refs.
\cite{Rempe1, Kimble1, Brennecke2007}. 
A different approach, which is also the one followed in this work, involves increasing the number of particles inside the cavity mode and studying
the collective strong coupling regime. For N atoms in the cavity mode the 
coupling scales as $g_{coll}=\sqrt{N} g$ \cite{TavisCummings1968}, 
the vacuum Rabi splitting scales as \cite{Transmission1}: 
\[
\Omega=2\sqrt{{{g_{coll}}^2-\left(\frac{\gamma-\kappa}{2}\right)^2}}
\]

and the collective 
cooperativity becomes
\[
C_{Coll}=\frac{N g^2}{2\kappa \gamma}
\]
The collective coupling of multiple atoms can increase the interaction 
dramatically. It has been observed in several experiments 
\cite{Raizen1989,Kasevich2006,Klinner2006,Brennecke2007}. In most experiments 
(except e.g \cite{Vuletic2003,BadCavity1,BadCavity2}) the coupling constant 
$g$ for the single atom is at least comparable to $\kappa$. 
In contrast, our aim is to investigate the coherent interaction between 
multiple cold atoms and a cavity mode in a lossy optical resonator, where the
single atom-cavity interaction is governed by the incoherent atomic decay rate
into free space. The large number of atoms only ensures that our system 
becomes strongly coupled.\\

In this work  we describe our experimental apparatus and present experimental results for 
the normal mode splitting of the cavity-atom system, a signature of  the strong coupling regime. 
Collective cooperativity parameters up to $C_{coll}>180$ are measured in our experiments.\\ 
A cooperativity parameter much larger than unity will allow us to study 
collective phenomena in atom-cavity systems, from cavity-assisted cooling 
of large samples of atoms \cite{Vuletic2003}, a model system for the 
cooling of molecules, to superradiance into the cavity mode and collective 
self-organisation of atoms.

\section{Experimental setup}

\begin{figure}[h!]
	\centering
	\includegraphics[width=0.25\textwidth]{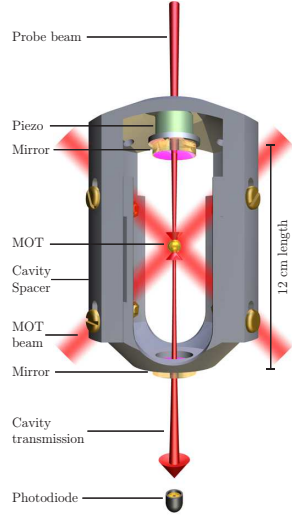}
	\caption{\label{fig:CavityPump} A schematic of the cavity arrangement and the MOT. The third pair of MOT beams is orthogonal to the image plane.}
\end{figure}

In the experiment we load a Magneto-Optical Trap (MOT) of Cesium atoms from the background gas directly into the mode of a nearly confocal optical resonator. The cavity and the MOT are depicted in figure \ref{fig:CavityPump}.\\
The FWHM linewidth of the cavities $TEM_{00}$ mode $\nu_C$ was measured to be 
$2\pi\times\left(1.60\pm0.05\right)$ MHz. For a Free Spectral Range (FSR)
of $2\pi\times\left(1249.5\pm0.4\right)$ MHz this leads to a finesse of $F=\left(780\pm15\right)$.\\ With about 12 cm cavity length the single mode-single atom coupling constant $g$ is very small, smaller than the cavity half loss rate $\kappa$ and much smaller 
than the atomic excited state half decay rate $\gamma$. The cavity-atom interaction 
at the single atom level is completely governed by the decay into free space and 
the influence of the resonator can be treated as a perturbation. This regime is the 
so called "bad cavity limit". The relevant parameters for our system  are:
$\left(g,\kappa, \gamma\right)=2\pi\times\left(0.12,0.8,2.6\right)$ MHz.\\
To increase the coupling as well as the overall mode volume the cavity in the 
experiment was chosen to be confocal (the length of the resonator equals the 
curvature radius of the mirrors). In this ideal case the higher order eigenmodes of 
the cavity are degenerate in frequency with the TEM$_{00}$-order mode, so that the 
coupling to each mode should contribute to the overall interaction.\\

We notice that the overlap of many transverse modes with slightly different frequencies leads to an empty-cavity linewidth larger than the $TEM_{00}$ value of $2\pi\times\left(1.60\pm0.05\right)$ MHz. The degeneracy is lifted due to abberations on the mirrors and a small deviation from the confocality condition. To measure the TEM$_{00}$-order linewidth the contributions of the higher order transverse modes were spatially filtered out 
by a small aperture. For the observation of the vacuum Rabi splitting the filter was then removed. The transmission now contains higher order transversal mode components and appears broader.\\
\\
Two important parameters need to be controlled separately to study the behaviour of the 
cloud in the mode with frequency $\omega_C$ illuminated by a pump beam with frequency 
$\omega_P$. First the cavity pump laser - atomic transition detuning $\Delta_A=\omega_P - \omega_A$  needs to be adjustable and stable on long time scales to enable averaging of several experimental runs. This is done by a home built extended cavity diode laser (ECDL) offset locked to the repumper laser of the MOT. The offset lock \cite{Weitz2004} 
enables us to gain atomic transition stability at an adjustable detuning  
$\Delta_A= 2\pi\times\left(-3 \rightarrow +2\right)$ GHz from the Cs $D_2$ $F=4 \rightarrow F'=5$ 
 transition. The other important experimental parameter is the cavity pump laser - cavity resonance detuning $\Delta_C=\omega_P - \omega_C$. Since there are cavity resonances every free spectral range the maximum detuning is $\Delta_C=\pm \left(FSR/2= 2\pi \times624.25\right) MHz$. The cavity resonance stabilization scheme is described in the next section.

\subsection{Cavity stabilization}

\begin{figure}[h]
	\centering
		\includegraphics[width=0.9\textwidth]{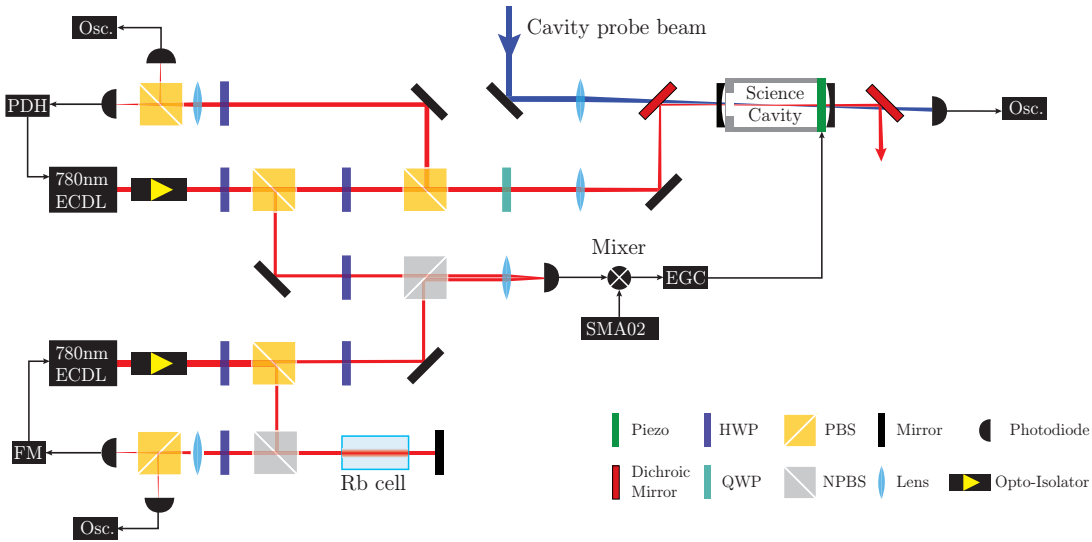}
	\caption{\label{fig:CavityStabi} The schematic of the experimental setup to control the cavity frequency. The beat signal of the two ECDL is mixed down with a function generator from R\&S (SMA02A). The error signal to stabilize the cavity piezo is generated with an error signal generating circuit (EGC) following the design of Ref. \cite{Weitz2004}.}
\end{figure}

During each experiment it was crucial that the cavity kept a stable, 
constant resonance frequency $\omega_C$, both with respect to the atomic transition 
frequency $\omega_A$, and with respect to the probe laser $\omega_P$. We established 
that by actively stabilizing the cavity to an atomic rubidium resonance. In order to be able to change the cavity resonance frequency, the locking point needed to be frequency offset by an adjustable radio frequency between 500 and 1500 
MHz provided by a radio-frequency generator. The stabilization schematic can be seen 
in figure \ref{fig:CavityStabi}.

Because of widespread availability we use a 780nm Extended Cavity Diode Lasers (ECDL) \cite{ECDL1995} stabilized via FM-Spectroscopy to the Rubidium $F=2\rightarrow F'=2,3$ crossover line as an atomic reference signal. Another ECDL lasing at a similar wavelength is injected into the cavity on a high-order TEM mode (a so called v-mode). This is possible in nearly confocal cavities and has the benefit of very small mode overlap with the MOT. No influence of this laser onto the cloud could be detected. It is stabilized to the mode using a Pound-Drever-Hall lock (PDH) \cite{PDH2001} and constantly kept on resonance to it.\\ It is basically measuring the length of the cavity very precisely. When the cavity length changes, the cavity resonance $\omega_C$ moves and the laser follows. To gain atomic transition stability for the cavity length this laser is now offset locked via a side-of-filter \cite{Weitz2004} technique to the 780 nm reference laser. The error signal is fed back to drive the piezo of the cavity, which can change the length of the cavity by about $17\mu m$, compensating mostly for slow temperature drifts and low frequency acoustic noise. The offset lock enables us to change the frequency difference of the two ECDLs so that we can position a cavity resonance at arbitrary detunings to both the atomic Cesium transition as well as the cavity probe laser. The lock is stable for the whole day with small drifts well below one cavity linewidth. The overall stabilization bandwidth is limited by the mechanical resonance of the science cavity spacer to about 800Hz.\\

\section{Strong Coupling in the bad cavity limit}
	One experimental cycle involved loading the MOT for an arbitrary length of time (between 100 and 3500ms), turning off the MOT cooling beams and the magnetic field followed by pumping the cavity with a weak probe beam. After a short off time, in which the cavity probe laser or the cavity could be moved in frequency the cycle started again. 
	The atom number interacting with the mode was controlled by choosing different MOT loading times.
	Three different experimental procedures were implemented to observe the normal mode splitting of the cavity-atom system. Two methods involved observing the transmission of the cavity as a function of the probe laser frequency. The third method relied on the collection of the scattered light into free space with a CCD camera during the probe. The observation of the Rabi splitting in the MOT fluorescence confirmed the results, but the data presented in this publication originates solely from transmission measurements.\\
 
 \begin{figure}
\begin{center}
\subfigure[]{
\resizebox*{6.5cm}{!}{\includegraphics{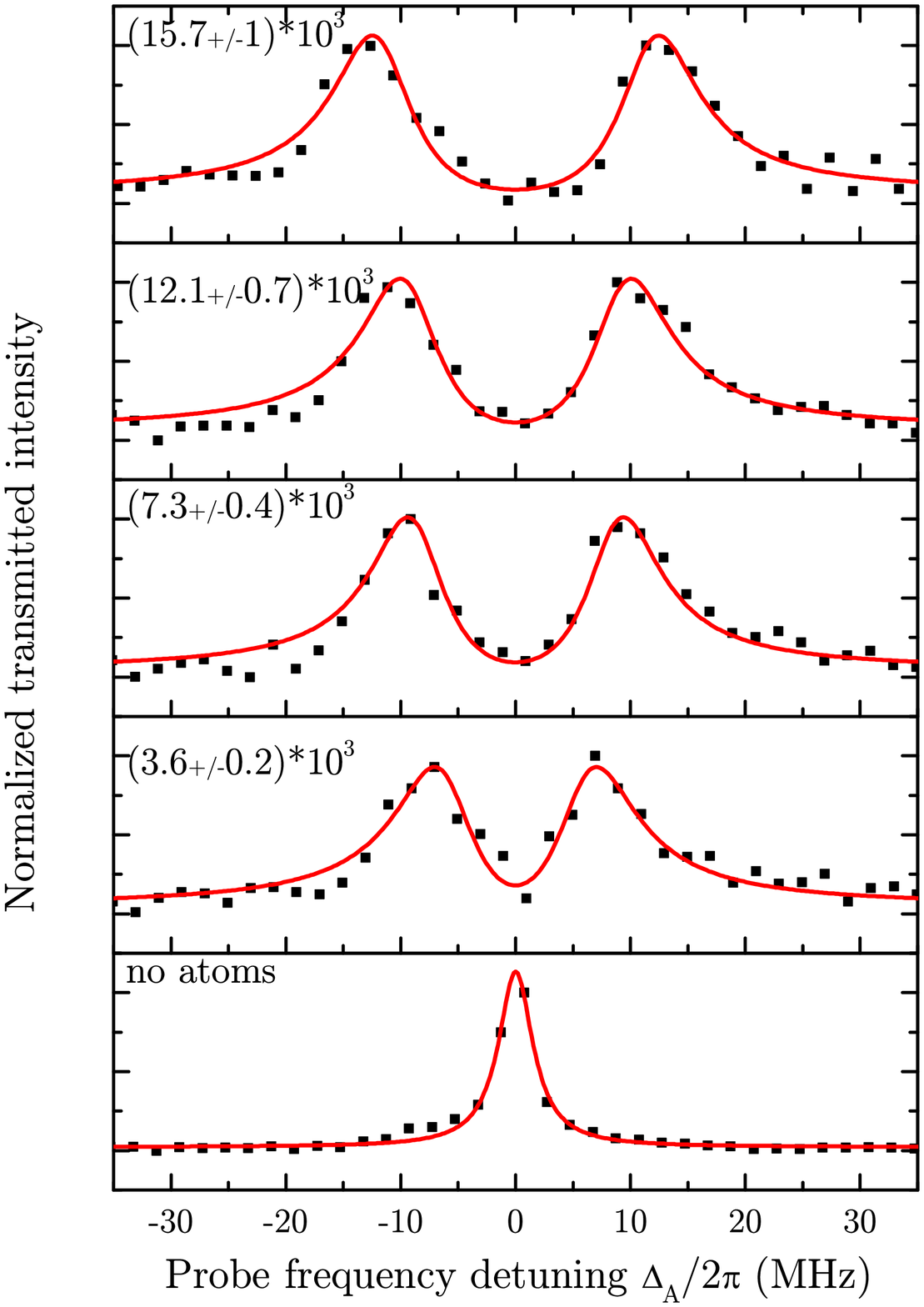}}}%
\subfigure[]{
\resizebox*{6.5cm}{!}{\includegraphics{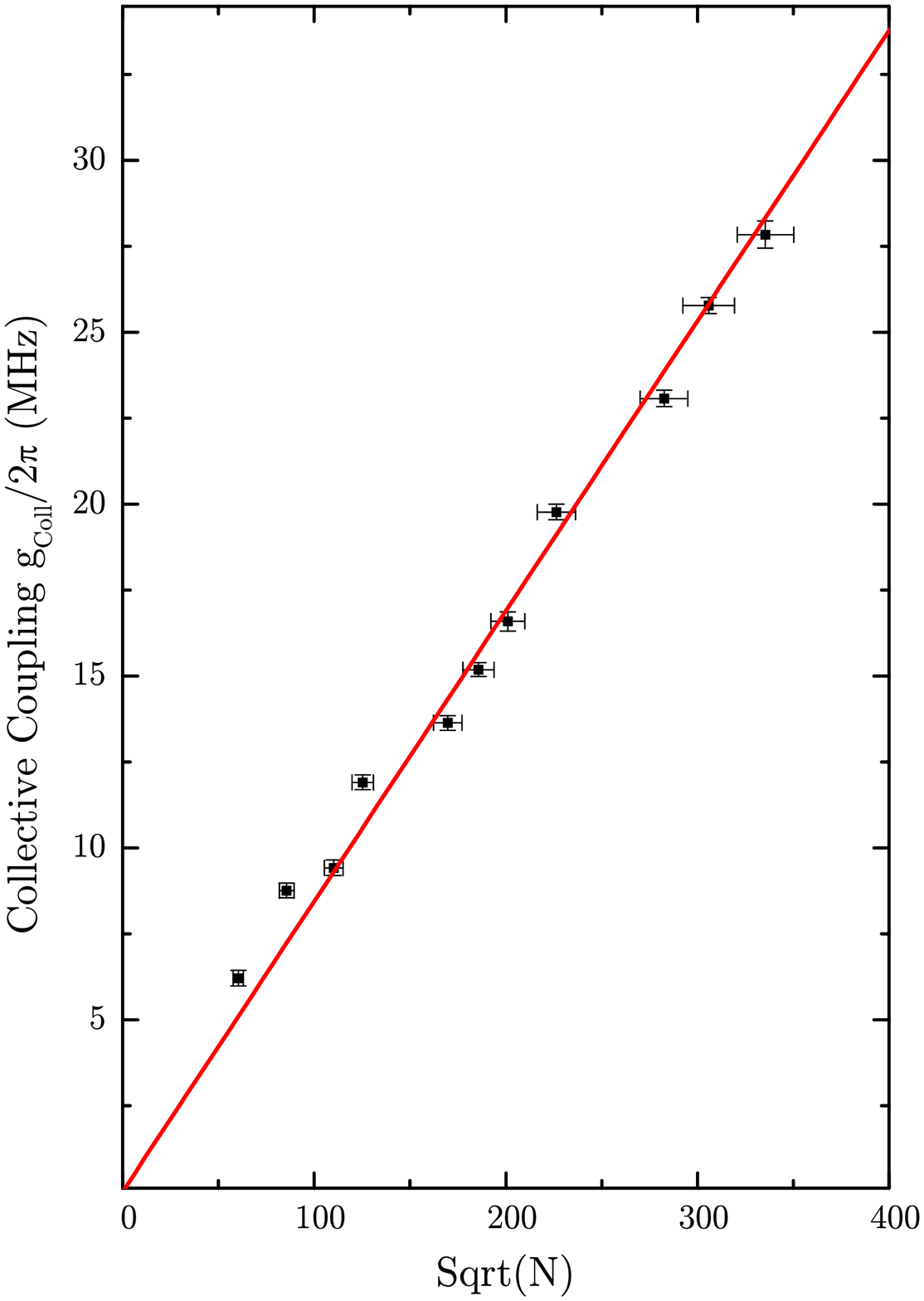}}}%
\caption{\label{fig2_a} a) Cavity transmission measurements with a weak incident $\sigma_+$-probe beam for different atom numbers in the mode. The cavity is positioned at the Cs $D_2$ $F=4 \rightarrow F'=5$ transition. The red curves are fits with the cavity transmission function from \cite{Transmission1} with free parameters $\kappa$ and $g_{coll}$. The scatter of the data can be explained by shot-to-shot atom number fluctuations as well as probe and cavity resonance frequency drifts. A conservative estimate for the atom number fluctuations is 6\% and $2\pi\times0.3$ MHz for the maximum drift of cavity resonance frequency and probe beam frequency during the whole experiment. b) The collective coupling constant displays the typical $\sqrt{N}$-behaviour indicating the strong collective coupling regime with an effective coupling parameter $g_{eff}=\textbf{$g$}/\sqrt{2}$. The linear fit to the data was used to calibrate the atom number in the mode.}%
\label{sample-figure1}
\end{center}
\end{figure}
The first set of transmission measurements used a constant probe laser frequency for
each point. The MOT was loaded for a certain time to a specific density, then 
switched off and a weak cavity probe beam was turned on with a 
mechanical shutter about $250\mu s$ later. The transmitted peak power for most experiments was measured to be below $\left(2.0\pm0.2\right)$ nW. According to Ref. \cite{transmission2}, this is much smaller than the critical photon number to cause bistable behaviour and therefore suitable to observe stable Rabi splitting. The shutter was left open for two milliseconds and we averaged the transmission signal on the photodiode over 1 ms. After 
that, the cavity probe laser was moved in frequency by $2\pi\times2$ MHz and the cycle started 
again. The scan of the probe beam begins at a detuning $\Delta_A=-2\pi\times 50$ MHz and ends at  
$\Delta_A=2\pi\times 50$ MHz so that for each curve 51 points were taken. The 
cavity resonance was tuned so to have the same frequency as the atomic transition, so that $\Delta_A-\Delta_C=\omega_C-\omega_A=0$. For a rough alignment the empty cavity transmission was compared to a Doppler-free Cesium spectroscopy while scanning. 
Then the cavity was locked and a first transmission measurement taken. If there 
was an amplitude difference in the two peaks indicating a detuning between atomic 
and cavity resonance frequency, the cavity was moved with the offset lock until 
the transmission peaks were similar in strength.\\ 
Figure \ref{fig2_a} shows data taken for the cavity positioned at the 
Cs $D_2$ $F=4\rightarrow F'=5$ transition. With respect to the empty cavity resonance the doublet 
splits with increasing number of atoms according to the $\sqrt{N}$ behaviour. 
The aquired data was fitted with the theoretical curve for the cavity transmission in the linear regime (see e.g Ref. \cite{Transmission1}) with the collective coupling $g_{coll}$ and the cavity half decay rate $\kappa$ as free parameters. The collective coupling is then displayed as a function of the square root of the atom number in figure \ref{fig2_a} b. The fit result for the cavity half loss rate is $\kappa_{fit}=2\pi\times\left(5.8\pm0.2\right)$ MHz for all the data and is therefore about 7 times larger than the measured $\kappa$ of the TEM$_{00}$ mode. We attribute this fact to the multimode structure of the nearly confocal resonator: higher order TEM modes have a bigger mode volume and therefore a smaller coupling constant g. This causes the higher order normal mode splitting to be smaller which appears as broadening of $\kappa$.
To calibrate the atom number in the mode we plotted the collective coupling constant over the square root of the observed MOT fluorescence first.  The MOT fluorescences was measured with a CCD camera for the same experimental parameters but without probing the cavity mode for each loading time. The shot to shot fluctuations are estimated to be below 6\%.  
The result was fitted with a line through the origin and the horizontal axis of figure \ref{fig2_a} b rescaled in a way that the gradient corresponded to the effective coupling constant $g_{eff}= g/\sqrt{2}=2\pi \times 85 kHz$. Since the probe is too weak to trap the atoms they see an averaged coupling during their ballistic movement in the 1 ms probe time, being zero at the nodes and $g$ at the antinodes of the cavity field. This results in the smaller coupling constant 
$g_{eff}$ according to Ref. \cite{Leslie2004}. The largest observed collective coupling of 
$g_{coll}=2\pi\times \left(27.8\pm0.4\right)$ MHz corresponded to $\left(1.33\pm0.08\right)\times 10^5$ atoms effectively coupled to the cavity mode and a maximum collective cooperativity of $C_{coll}=186\pm5$. The rescaling factor relating the atom number in the MOT to the atom number coupled to the mode was 1.47. It is bigger than 1 because not all atoms in the MOT are in the center of the mode. 


The collective behaviour of the vacuum Rabi splitting of the cavity-atom system is clearly demonstrated. The coupling to the resonator dominates the environmental dissipation process indicating the collective strong coupling regime for atom numbers in the mode bigger than about 600 corresponding to $C_{coll}=1$.\\
\begin{figure}
\begin{center}
\subfigure[]{
\resizebox*{6.5cm}{!}{\includegraphics{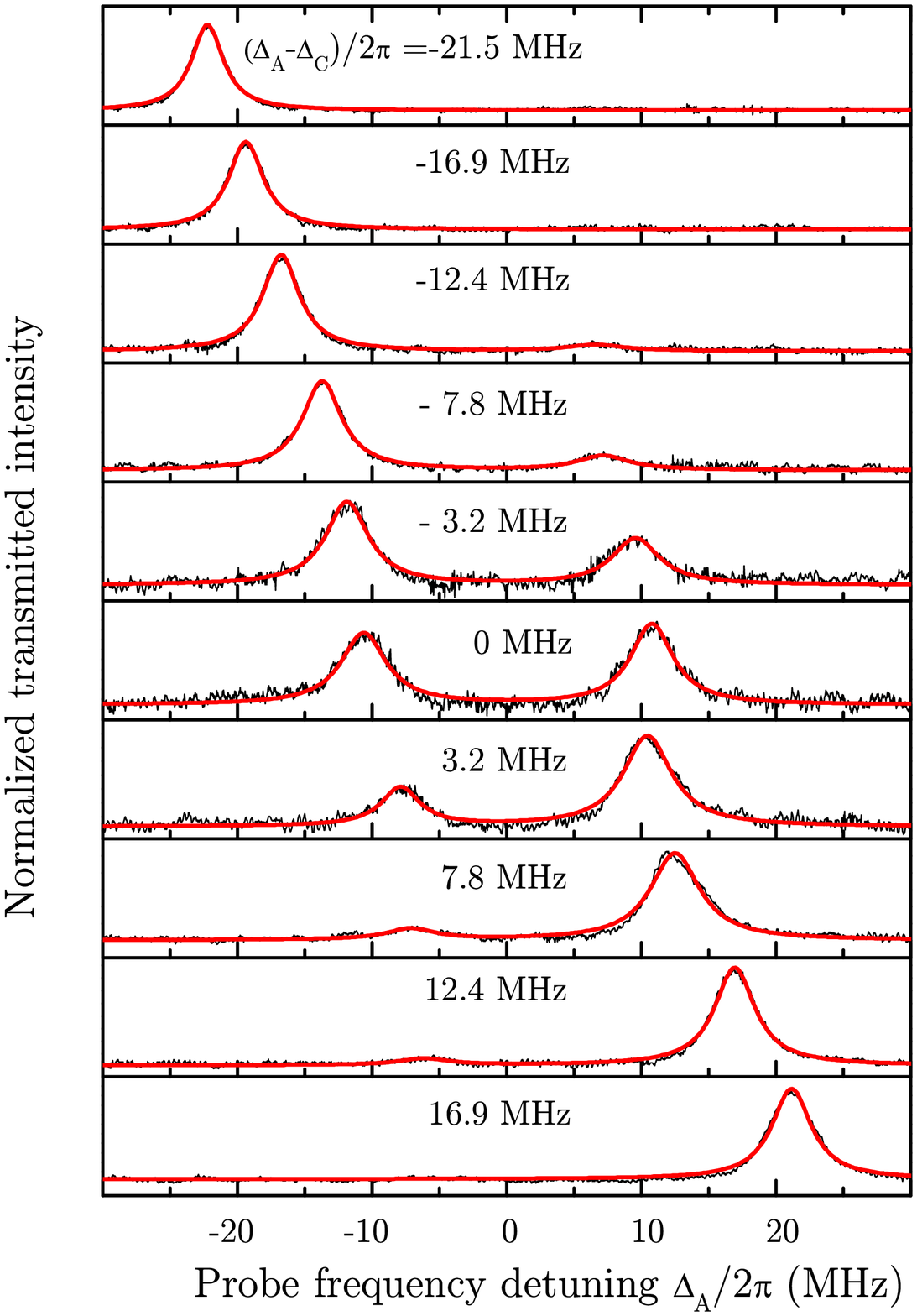}}}%
\subfigure[]{
\resizebox*{6.5cm}{!}{\includegraphics{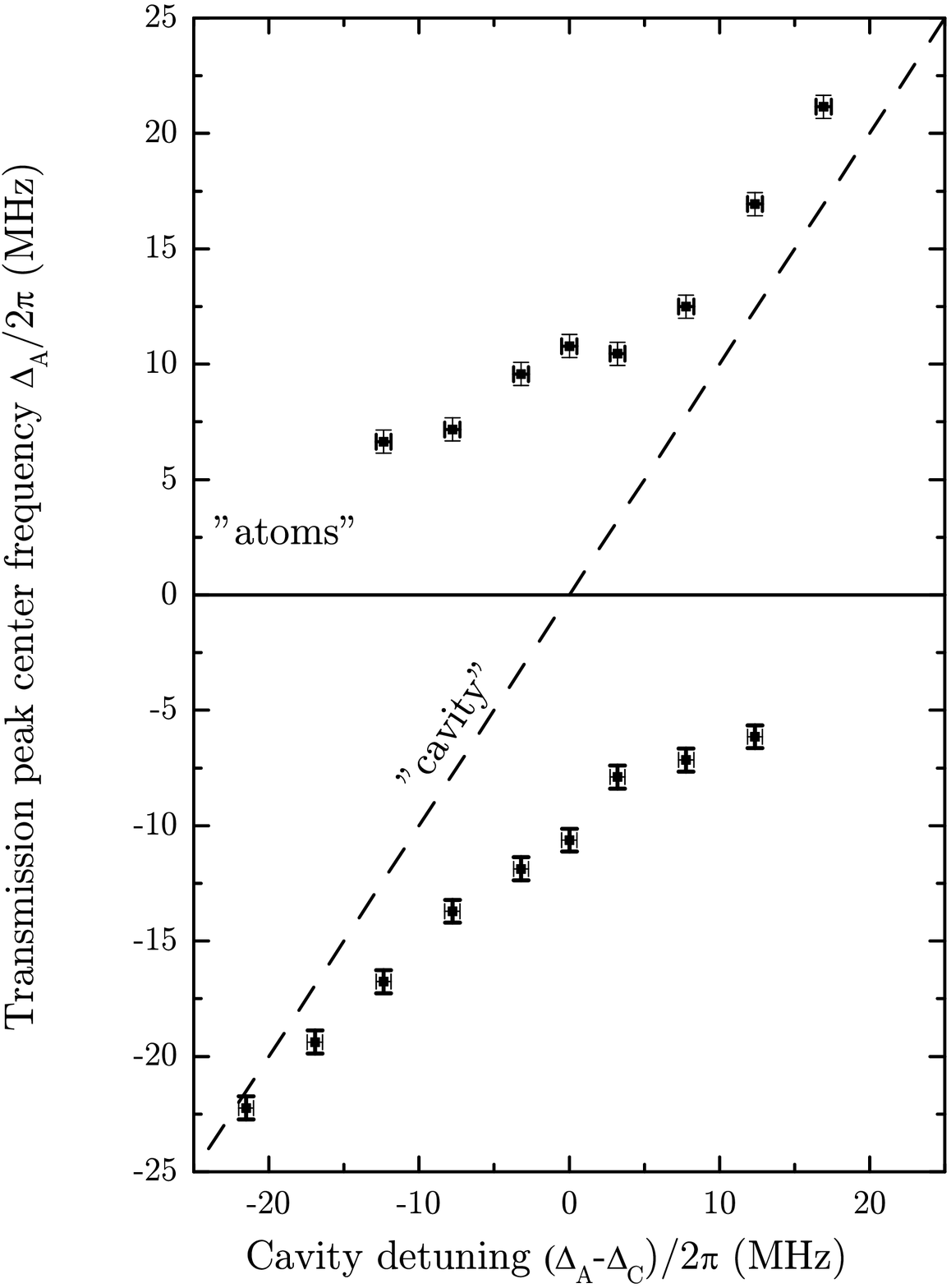}}}%
\caption{\label{fig2} a) The transmission of $\sigma_+$-polarized light through the cavity as a function of the probe laser frequency for an atom number $N=\left(15\pm22\right)\times10^3$. The different traces (twenty averages each) are taken for different cavity resonance-atomic transition detuning. The error in the detuning is estimated to be below $2\pi\times0.3$ MHz. The difference between each trace is about $2\pi\times4.6$ MHz (except the middle three, where the frequency difference was $2\pi\times3.2$ MHz) starting from the top at $-2\pi\times21.5$ MHz. The red line is a double lorenzian fit to those curves where two peaks were observable. b) The center frequencies of the two peaks displayed over the cavity resonance - atomic transition detuning. For an empty cavity the mode position should follow the dashed line. Coupling atoms to the cavity splits the resonance and an avoided crossing can be observed.}%
\label{sample-figure}
\end{center}
\end{figure}
The second set of measurements was taken in a slightly different way. Instead of pumping the mode for a certain time with the same frequency after the release from the MOT, we scanned the probe laser over the atomic transition. The scan speed was about 40 MHz/ms. Eventhough the ballistic expansion of the cloud during the first 2 ms is negligible, the two peaks are probed one after the other. For high intensities of the probe light, this causes the second peak to become asymmetric in height and position even if the empty cavity resonance equals the atomic transition. For small intensities this was not observable.
The traces of the transmission were recorded with an oscilloscope and averaged over 20 shots. They can be seen in figure \ref{fig2} a. After aquiring the waveform the cavity offset lock was used to move the cavity resonance with respect to the atomic transition. In this way the avoided crossing induced by the collective cavity-atom coupling could be observed. Figure \ref{fig2} b shows how, for large cavity-transition detunings, the coupling is small, therefore the eigenfrequency of the cavity mode with atoms does not deviate from the case without atoms (indicated by the dashed line). Approaching the atomic resonance causes the coupling to increase and the mode to split.\\ 
  
\begin{figure}
\begin{center}
\subfigure[]{
\resizebox*{6.5cm}{!}{\includegraphics{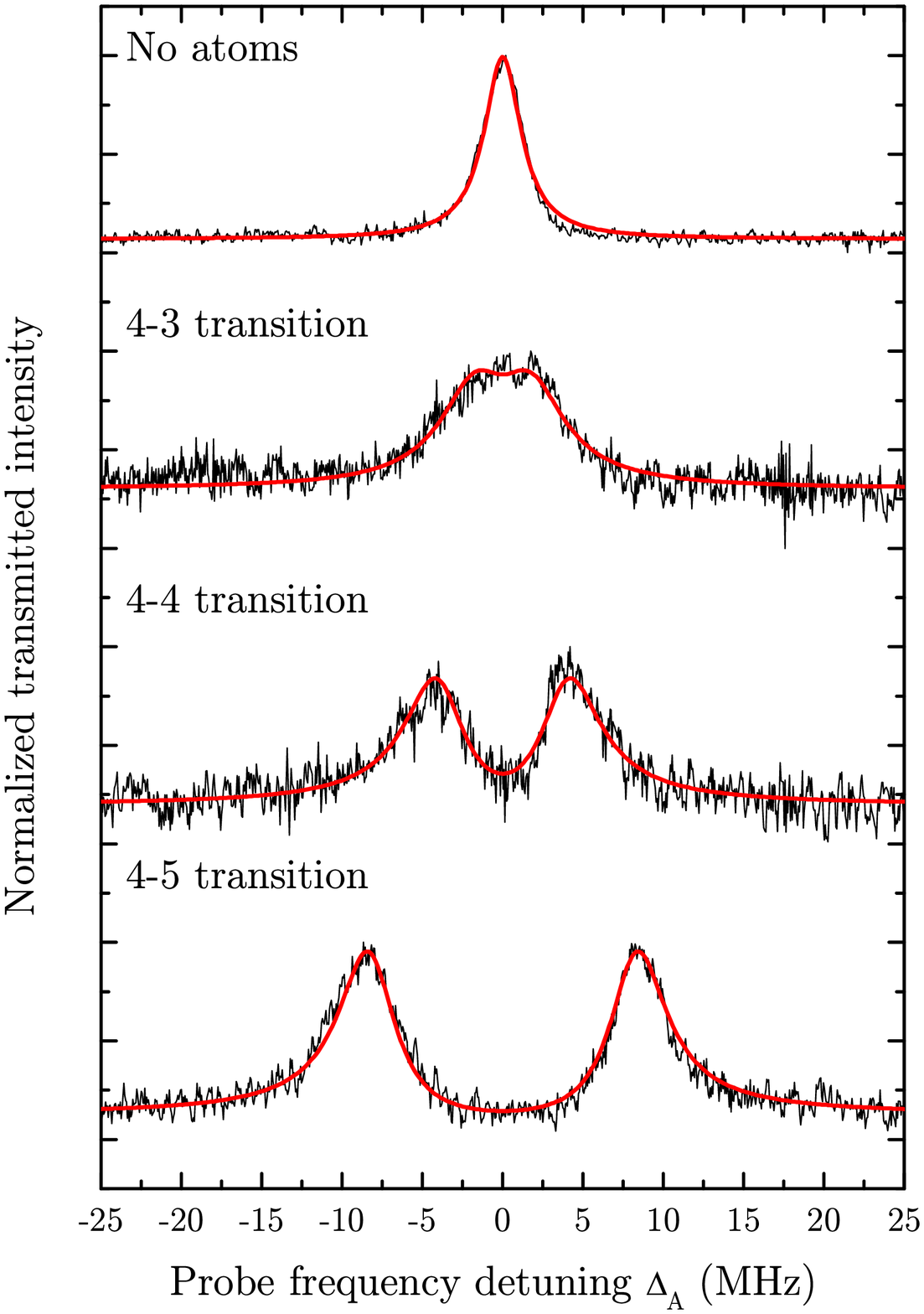}}}%
\subfigure[]{
\resizebox*{6.5cm}{!}{\includegraphics{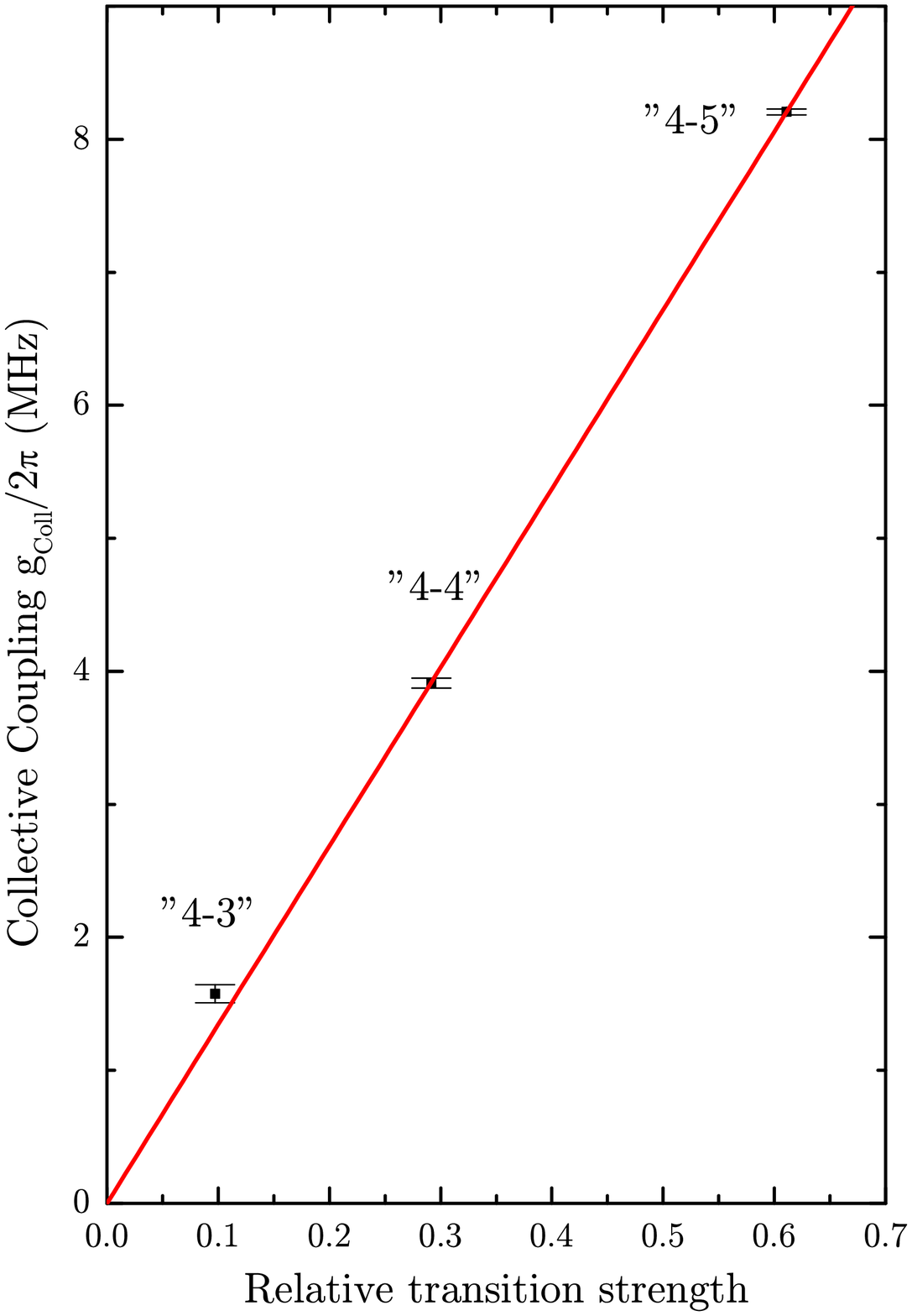}}}%
\caption{\label{fig3} a) The transmission of the cavity of a weak $\pi$-polarized probe beam for each hyperfine transition in comparison with the empty cavity. The atom number in the cavity is $N=\left(110\pm7\right)\times 10^3$. The stronger the dipole moment of the transition the bigger the splitting. The red line is the theoretical curve for the transmission according to Ref. \cite{Transmission1}. b) The observed collective coupling constant as a function of the relative transition strength as given in Ref. \cite{Steck}.}
\label{sample-figure}
\end{center}
\end{figure}

The last set of measurements involved observing the splitting for the other hyperfine states of cesium as seen in figure \ref{fig3} a. The cavity was positioned according to the procedure for the 4-5 transition. The MOT loading time and all other parameters are the same for each transmission measurement but the observed splitting is less. 
Our data shows that the splitting is proportional to the relative hyperfine transition 
strength, consistent with the fact  that the  coupling constant $g$ is proportional 
to the relative transition strength.\\
\section{Conclusion}
In conclusion we have demonstrated the strong coupling of an ensemble of up to $\left(1.33\pm0.08\right) \times 10^5$ atoms with a lossy nearly confocal cavity leading to a collective 
cooperativity parameter of up to $C_{coll}=186\pm5$. In contrast to other observations of the avoided crossing and the behaviour of the vacuum Rabi splitting (e.g \cite{Raizen1989, Kasevich2006}) the single atom - single mode coupling $g$ is very small compared to $\kappa$ and the relevant excited state half decay rate $\gamma$. The large number of inner cavity scatterers only enables us to reach the strong coupling regime which is a prerequisite to study certain cavity cooling mechanisms. This system is also suitable for the study of other collective effects such as superradiance or collective selforganization in a lossy cavity.\\

\section{Acknowledgement} This work is supported by the EPSRC grant EP/H049231/1. We would like to thank Jon Goldwin for stimulating discussions.\\

\bigskip

\bibliographystyle{tMOP}
\bibliography{bibi1}

\label{lastpage}

\end{document}